\input epsf
\documentstyle[preprint,aps]{revtex}
\begin{document}

\title{Ill-posedness of a double null free evolution scheme for black
hole spacetimes}

\author{Carsten Gundlach}

\address{LAEFF-INTA (Laboratorio de Astrof\'\i sica Espacial y F\'\i sica
Fundamental -- Instituto Nacional de Tecnolog\'\i a Aerospacial) \\
PO Box 50727, 28080 Madrid, Spain}

\author{Jorge Pullin}

\address{Center for Gravitational Physics and Geometry, Department of
Physics, 104 Davey Lab\\The Pennsylvania State University, University 
Park, PA 16802}

\date{June 10, revised October 24, 1996}
\maketitle

\maketitle

\begin{abstract}
We suggest that ``free evolution'' integration schemes for the
Einstein equations (that do not enforce constraints) may contain
exponentially growing modes that render them useless in numerical
integrations of black hole spacetimes, independently of how the
equations are differenced. As an example we consider the evolution of
Schwarzschild and Reissner-Nordstr\"om 
spacetimes in double null coordinates.
\end{abstract}

\section{Introduction}

There are two steps from the Einstein equations to a numerical code
for solving them: First one selects the fields of the problem (the
metric, its derivatives, connection coefficients etc.) and a subset of
the Einstein equations to evolve these from some boundary
conditions. This constitutes the ``differential problem''. The
resulting set of differential equations is then transformed into a set
of difference equations on a numerical grid, constituting the
``difference problem''. (Here the term ``boundaries'' includes
spacelike and null as well as timelike boundaries.)

In the first step, one distinguishes between ``free'' and
``constrained'' evolution schemes.  In a free evolution scheme, the
constraints are imposed only at the boundaries, and all fields are
propagated by means of the evolution equations. In a constrained
scheme, one uses fewer evolution equations, and instead reconstructs
some fields at each time step from the constraints. (Here the term
``time'' refers to the coordinate that labels slices in the numerical
evolution, which can be null. Constraints are all equations restricted
to a hypersurface.)

In this paper we show that free evolution may already be ill-posed as
a differential problem in some cases because it admits exponentially
growing perturbations which are solutions of the evolution equations,
but which violate the constraints and are therefore unphysical. For
analytic purposes this does not matter, as imposing the constraints on
the boundaries assures that they are obeyed everywhere, but any
finite-differencing scheme introduces small perturbations which do not
generically obey the constraints, and serve as initial data for the
growing unphysical modes. The amplitude of these modes depends
therefore on the choice of difference scheme (and decreases with the
grid spacing), but their evolution (for example, the rate of
exponential blow-up) is determined by the differential problem alone.

As an example differential problem we consider free evolution in
spherical symmetry in double null coordinates. It is sufficiently
simple that we can calculate the blowup of the unstable mode
analytically, before confirming it in a testbed calculation.  This
particular instability is a serious problem inside and just outside
black holes, harmless far outside black holes, and is not present in
perturbations around flat spacetime.

\section{Double null coordinates in spherical symmetry}

The metric of any spherically symmetric spacetime can be written as 
\begin{equation}
\label{metric}
ds^2 = - 4 f \, du \, dv + r^2 \, d\Omega^2,
\end{equation}
where the metric coefficients $f$ and $r$ depend on $u$ and $v$ only.
Our matter will be a minimally coupled massless real scalar field $\phi$,
plus a spherically symmetric Maxwell field. The Maxwell field has no
sources, and is therefore constrained to be a pure Coulomb field of
constant charge $q$ anchored at (the singularity) $r=0$.

This model has been used in both analytic \cite{Israel} and numerical
\cite{Gnedin,Brady} work as a toy model for studying the interior of
realistic black holes. Realistic black holes should be spinning, at least
slightly, and would therefore be expected to have a Cauchy horizon, like
the Kerr solution. On the other hand they would be inevitably perturbed by
gravitational wave tails, and these perturbations would blow up on the
Cauchy horizon. What really happens has therefore been the subject of
prolonged investigation. Replacing the rotation by an electric charge and
gravitational waves by the scalar field allows the simplification of
spherical symmetry. Although we have chosen a particular matter model here,
we shall see that the instability arises in the gravitational part of the
equations and is therefore independent of the matter model.
 
The $uu$ and $vv$ components of the Einstein equations are
\begin{eqnarray}
&& r_{,uu} - {f_{,u} \over f} \, r_{,u} + \left(\phi_{,u}\right)^2
\equiv E_1 = 0, \\
&& r_{,vv} - {f_{,v} \over f} \, r_{,v} + \left(\phi_{,v}\right)^2
\equiv E_2 = 0.
\end{eqnarray}
The $\theta\theta$
component is
\begin{equation}
r_{,uv} + {r_{,u} r_{,v} \over r} - {f \over r} + {q^2 f \over r^3} 
\equiv H_1 = 0,
\end{equation}
and the $uv$ component is 
\begin{equation}
\label{H2}
\left(\ln f\right)_{,uv} + 2 {r_{,uv} \over r} - 2 {q^2 f \over r^4} 
+ 2 \phi_{,u} \phi_{,v} \equiv H_2 = 0.
\end{equation}
The other components of the Einstein equations are redundant.
The massless wave equation for $\phi$, restricted to spherical
symmetry, is 
\begin{equation}
\phi_{,uv} + {r_{,v} \over r} \phi_{,u} + {r_{,u} \over r} \phi_{,v}
\equiv H_3 = 0.
\end{equation}
We then have three hyperbolic equations $H_1$, $H_2$ and $H_3$ and two
elliptic equations $E_1$ and $E_2$. (The elliptic equations are really
ordinary differential equations because of the spherical symmetry.)
The elliptic equations are propagated by the hyperbolic equations in
the sense of
\begin{equation}
E_{1,v} = H_{1,u} + 2 r \phi_{,u} H_3 - {r_{,v} \over r} E_1 
- r_{,u} H_2 + \left( {r_{,u} \over r} - {f_{,u} \over f} \right) H_1
= 0,
\end{equation}
while a similar equation holds for $E_{2,u}$.

\section{The free evolution scheme}

A possible evolution scheme, perhaps the most natural one, and certainly
one easy to implement numerically, is to consider the three hyperbolic
equations as evolution equations for $f$, $r$ and $\phi$, and the elliptic
equations as constraints which are imposed only on the boundary. The
natural initial value problem for these equations, is a double null initial
value problem. $r$, $f$ and $\phi$ are given as functions of $u$ on the
null cone $v=v_0$, subject to the constraint $E_1=0$, and as functions of
$v$ on the intersecting null cone $u=u_0$, subject to the constraint
$E_2=0$. At the intersection point $(u=u_0, v=v_0)$ it is sufficient that
the data $r$, $f$ and $\phi$ be continuous. In the free evolution scheme,
the constraints are imposed only on the null boundary data, but are not
used for the numerical solution inside the numerical domain. 

We now examine the stability of this free evolution scheme in the context
of black hole physics. As a testbed case we use the Reissner-Nordstr\"om
solution. It is the unique solution for $\phi=0$, and it is known in double
null coordinates in (essentially) closed form. In one particular gauge
choice (Eddington-Finkelstein coordinates) this solution is
\begin{equation}
\label{RN1}
f(u,v) = f(r) = -{1\over 4} \left( 1 + {2M\over r} - {q^2 \over r^2}
\right),
\end{equation}
where $r(u,v)$ is given implicitly by
\begin{equation}
\label{RN2}
r(u,v) = r(r_*), \quad r_*={v - u \over 2}, \quad {dr \over dr_*} = f(r).
\end{equation}
The implicit equation $dr/dr_*=f(r)$ can be solved numerically to arbitrary
precision so that the solution exists in closed form for all numerical
purposes.

In our test case we have set $\phi=0$ because we expect a
nonvanishing $\phi$ to give rise to a physical instability, namely
mass inflation, inside the black hole, while we want to check if such
instabilities are already contained in the free evolution scheme in a
situation when we know that no physical instabilities can be present.

In order to look for instabilities in analytic approximation, it is
useful to make a change of variables in order to eliminate first
derivatives in the hyperbolic equations. With the new variables
\begin{equation}
\label{new}
y \equiv \ln f + \ln r, \qquad x \equiv r^2
\end{equation}
the evolution equations $H_1=0$ and $H_2=0$, now restricted to
$\phi=0$, become
\begin{equation}
x_{,uv} + A(x,y) = 0, \qquad y_{,uv} + B(x,y) = 0.
\end{equation}
(The evolution equation $H_3=0$ for $\phi$ is dropped.)
The constraints $E_1=0$ and $E_2=0$ become
\begin{equation}
x_{,uu} - y_{,u} \, x_{,u} = 0, \qquad x_{,vv} - y_{,v} \, x_{,v} = 0.
\end{equation}
Now we linearize the evolution equations, denoting the perturbations of
$x$ and $y$ by $\xi$ and $\eta$:
\begin{equation}
\label{pert}
\xi_{,uv} + A_{,x} \xi + A_{,y} \eta = 0, \qquad
\eta_{,uv} + B_{,x} \xi + B_{,y} \eta = 0.
\end{equation}
The coefficients of these equations on the Reissner-Nordstr\"om
background are
\begin{equation}
A = A_{,y} = - 2 f \left(1 - {q^2 \over r^2} \right), \quad
B = B_{,y} = A_{,x} = {f \over r^2} \left(1 - 3 {q^2 \over r^2}
\right), \quad
B_{,x} = - {3\over 2} {f \over r^4} \left(1 - 5 {q^2 \over r^2}
\right).
\end{equation}
 
To obtain an analytic approximation to the perturbations we now
consider $r$ and $f$ of the background as slowly varying functions of
$u$ and $v$. We make a mode ansatz
\begin{equation}
\xi(u,v) = \xi_k e^{i \omega(v+u) + i k (v-u)}, \quad
\eta(u,v) = \eta_k e^{i \omega(v+u) + i k (v-u)}.
\end{equation}
The ansatz turns the derivative $\xi_{,uv}$ into the algebraic expression
$(k^2-\omega^2)\xi_k$, and the equations (\ref{pert}) into the local
dispersion relation $\omega^2(k) = k^2 + \lambda_\pm$. The $\lambda_\pm$
are the eigenvalues of the two-by-two matrix $[(A_{,x}, A_{,y}), (B_{,x},
B_{,y})]$. Their value on the Reissner-Nordstr\"om background is
\begin{equation}
\lambda_\pm = {f \over r^2} \left[ (1 - 3 \rho) \pm \sqrt{3 (1 -
\rho)(1 - 5\rho)} \right], \quad {\rm where} \quad \rho\equiv q^2/r^2,
\end{equation}
and the corresponding eigenvectors $(\xi_k,\eta_k)$ are
\begin{equation}
\label{eigvec}
\eta_{k\pm} = \pm \, {\sqrt{3} \over 2 r^2} \, \sqrt{1 - 5 \rho \over 1 -
\rho}\ \xi_{k\pm}.
\end{equation}

With $u$ and $v$ both increasing to the future. $u+v$ labels time, and
$u-v$ labels space. $u+v$ increases away from the null boundary data.
(Inside the black hole, evolution towards increasing $u+v$ is also
evolution towards the singularity.) Therefore we have an instability if
$\omega^2(k)$ is negative for any $k$, that is, for $\lambda < 0$. We see
that for $q=0$, $\lambda_\pm = (1\pm \sqrt{3}) r^{-2} f$. Although $f$
changes sign at the horizons, one of the $\lambda$ is always negative for
all $r$.  For $q \ne 0$ the situation is not changed drastically. There is
now a region where both $\lambda$ are positive, namely $0.74\, |q| < r <
|q|$. But this is only a narrow band inside the black hole, and globally
the free evolution scheme is still unstable.

The transformations that leave the form (\ref{metric}) of the metric
invariant are $u \to U(u)$ and $v \to V(v)$, where $U$ and $V$ are
arbitrary functions. The speed of the exponential blowup itself is
gauge-invariant, in the sense that $\lambda$ transforms correctly under the
coordinate transformations $u \to U(u)$ and $v \to V(v)$. We can see this
noting, from (\ref{metric}), that $f$ transforms in the same way as
$g_{,uv}$ for any scalar $g$, and that $\lambda$ is the product of $f$
times a scalar ($r$ is a scalar), and hence also transforms like $f$. Our
results would therefore also hold in Kruskal coordinates, or any other
double null coordinates.

We should stress again that the unstable perturbations we have
constructed are not tied to a particular numerical scheme.  They are
solutions to the hyperbolic part of the Einstein equations. They must
arise in any numerical algorithm implementing a free evolution scheme
in null coordinates, simply because they are excited by the
discretisation error of any numerical scheme.

Of course these unstable modes are not solutions of the constraints,
or elliptic part of the Einstein equations. We can explicitly verify
this by constructing the perturbations that do obey the
constraints. The linearized constraints are
\begin{equation} 
\xi_{,uu} - y_{,u} \, \xi_{,u} - x_{,u} \, \eta_{,u} = 0, \quad
\xi_{,vv} - y_{,v} \, \xi_{,v} - x_{,v} \, \eta_{,v} = 0.
\end{equation}
Because the Reissner-Nordstr\"om solution is unique, any perturbations
of (\ref{RN1},\ref{RN2}) obeying all the Einstein equations must be
infinitesimal coordinate transformations. Perturbatively we write $u \to u + \mu(u)$ and $v \to v +
\nu(v)$. The corresponding perturbative changes in the metric
variables are
\begin{equation} 
\xi = - x_{,u} \, \mu - x_{,v} \, \nu, \quad 
\eta = - y_{,u} \, \mu - y_{,v} \, \nu - d\mu/du - d\nu/dv,
\end{equation}
Clearly these obey the linear constraints and are bounded. The
unstable modes we have given do not obey the linear constraints and
blow up.

We first encountered this instability in a code modeled on that of
reference \cite{Gnedin}, when we found we were unable to recover the
Reissner-Nordstr\"om solution in a testbed. Having constructed the
unstable modes analytically, we could quantitatively verify their
presence in the code. (Once more it should be said that the
instability is connected to the free evolution scheme, not to any
particular numerical implementation.) As an example we numerically
construct a null diamond (that is, a square in $u$ and $v$) of the
Schwarzschild solution centered on $r=3M$. (We have chosen $r \sim 3M$
because there the instability is greatest outside the horizon.)
Fig. 1 shows the numerical setup. At $r=3M$ in the coordinates
(\ref{RN1},\ref{RN2}), $\lambda_+ \simeq - 0.025\,M^{-2}$. This
corresponds to a blowup of the unstable modes as $\exp(0.32 \, M^{-1}
t)$, where $t=(u+v)/2$ is the usual Schwarzschild time
coordinate. Numerically we find that the error grows as $\exp(0.24 \,
M^{-1} t)$ at $r=3M$. The discrepancy of the exponent may be due to
the approximation of constant $r$ and $f$ we have made in the analytic
calculation.

There is a second prediction of the analytic model which can be verified at
least qualitatively. From the form of the eigenvectors (\ref{eigvec}),
together with the definitions (\ref{new}) it follows that the relative
error in $f$ is generically of the same order as the relative error in $r$,
except for $r \simeq |q|$, where it will much greater (because $1-\rho \ll
1$). This feature is confirmed by numerical evolutions inside and outside
the horizon, for various values of $q$.

\section{Conclusions}

We have given a simple clean example of a numerical instability
arising from constraint violation. The amplitude of the unstable mode
depends on the discretisation scheme, but its growth rate does not. We
have estimated the growth rate analytically, and have confirmed it for
a particular discretisation.

The free evolution scheme we have described was used for the numerical
study of perturbed Reissner-Nordstr\"om black holes in reference
\cite{Gnedin}. As our investigation shows, this scheme is unable even
to evolve the exact Reissner-Nordstr\"om solution (setting the scalar
field to zero) because of an exponential instability eventually
crashing the code. It can therefore not be used for the study of the
physical instability triggered by small physical perturbations.  (This
is the physical instability \cite{Israel} predicted to lead to mass
inflation and the destruction of the Cauchy horizon.)  This casts a
shadow on the physical results obtained using this scheme.

Other published codes \cite{Brady,GP,GW,GPP,Garfinkle,HS} using a
double null numerical grid use fully constrained evolution.  The interior
of a charged black hole has been treated with a fully constrained evolution
scheme based on coordinates $u$ (retarded time) and $r$ (curvature radius),
but evolving them on a double null grid \cite{Brady}. This kind of
algorithm has been pioneered in \cite{GP}, and has been used extensively
since \cite{GW,GPP,Garfinkle}. A code actually based on double null
coordinates, as well as a null grid, was used in \cite{HS}, implementing 
a fully constrained scheme.  Here fully constrained means that one uses
the maximum number of equations containing only $v$-derivatives.

Another instance of an ill-posedness and instability arising from
constraint violation has been found for the nonlinear evolution of
weak gravitational waves on a flat background
\cite{Anninosetal}. The growth rate of the instability depended on the
discretisation scheme, however, which is puzzling in the light of our
results.

Generally our results suggest that free evolution schemes are
generally harder (or impossible, as in this case) to make stable than
fully constrained schemes. This is not in conflict with the statement
\cite{Choptuik} that if one if one evolves in free evolution to a
certain order in the grid spacing, the constraints are automatically
obeyed to that order. The numerical, constraint-violating, error may
go down as some power of decreasing step-size, but it can also be
growing exponentially with time. If this exponential growth is fast
enough, one will, in typical situations, be unable in practice to
compensate for the error at late times by using a finer numerical
grid.

Enforcing all the constraints at each time step can avoid a blowup
only when the solution to be calculated is itself insensitive to small
perturbations in the initial data, because then all rapidly growing
perturbations must be constraint violations. The solution itself
however may be highly sensitive to perturbations in the initial data,
and/or may blow up at a spacetime singularity. Examples of such
physical instabilities are mass inflation in the perturbed black holes
we discussed here, the threshold of black hole formation, or chaos in
the mixmaster universe \cite{Hobill,Berger}. Such examples may not be
strictly well-posed but are nevertheless interesting. Then one must
enforce the constraints to distinguish physical from unphysical
(constraint-violating) perturbations.

\acknowledgments

We wish to thank Beverly Berger and Wai-Mo Suen for correspondence,
and Abhay Ashtekar, Karel Kucha\v{r} and Richard Price for discussions.
CG would like to thank Nick Gnedin for helpful conversations. 
This work was supported in part by grants NSF PHY92-07225, NSF-PHY
9423950 and by research funds of the University of Utah and the
Pennsylvania State University, its Office for Minority Faculty
Development and the Eberly Family Research Fund at Penn State. JP also
acknowledges support from the Alfred P. Sloan Foundation through an
Alfred P. Sloan fellowship.

\begin{figure}
\epsfysize=10cm
\centerline{\epsffile{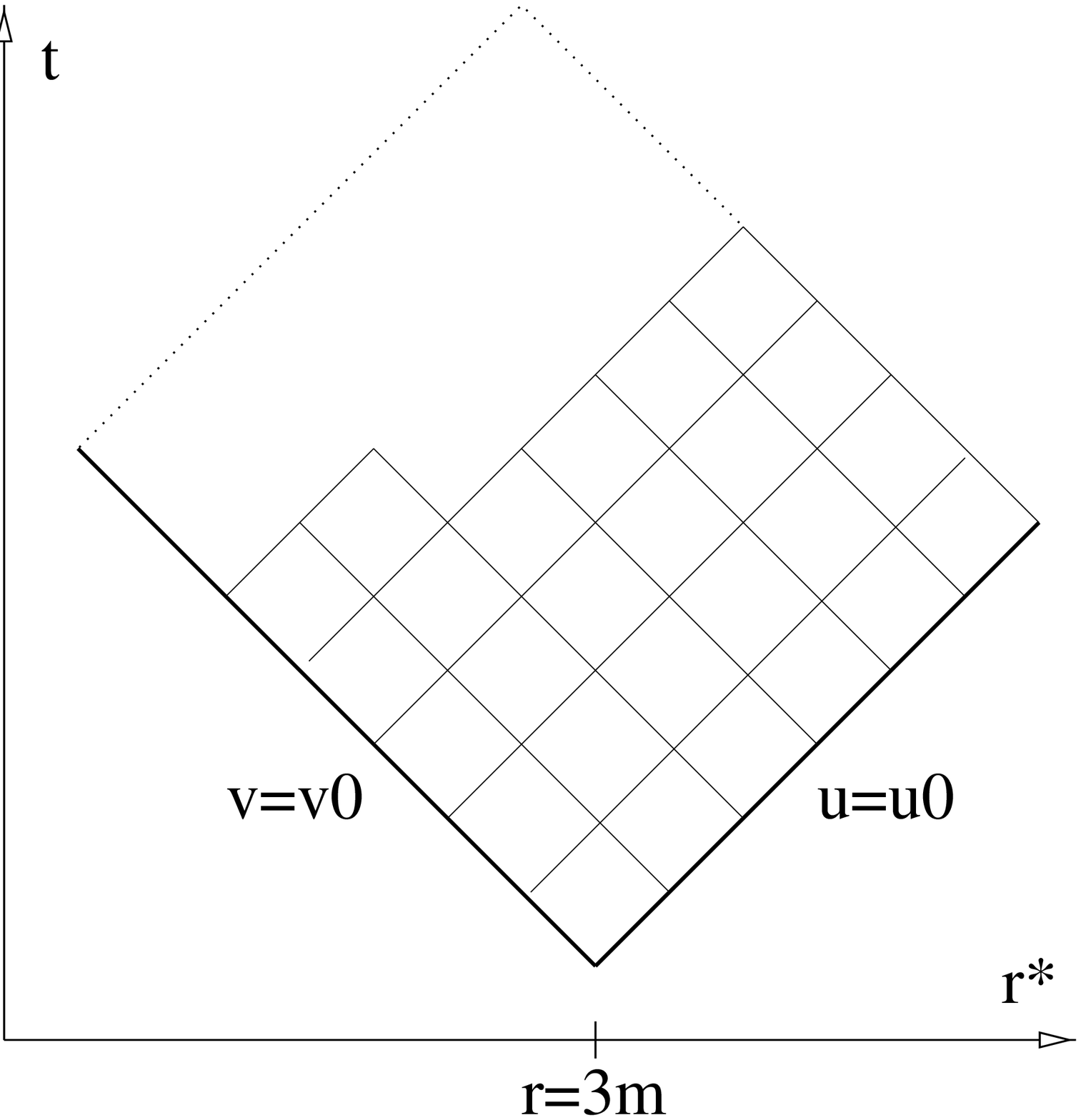}}
\caption{The numerical grid setup for our example. The partially
filled-in numerical grid indicates in what order grid points are
completed. The fat lines $u=u_0$ and $v=v_0$ are null boundaries. From
a numerical point of view one might call the former the initial time
and the latter a boundary, but clearly this is a matter of words.  The
dotted line marks the boundary of the domain of dependence of the null
boundary data.  $t$ and $r_*$ are the Schwarzschild time and tortoise
radial coordinate.}
\end{figure}

\end{document}